\documentclass[prb,twocolumn,showpacs]{revtex4}
\begin{document}
\title{High Temperature Mixed State $c-$Axis  Dissipation in Low
Carrier Density
$Y_{0.54}Pr_{0.46}Ba_{2}Cu_{3}O_{7-\delta}$}
\author{T. Katuwal}
\author{V. Sandu\thanks}
\thanks{Permanent Address: National
Institute of Materials Physics, 077125
Bucharest-Magurele, Romania}
\author{C. C. Almasan }
\affiliation{Kent State University, Kent, OH-44242 }
\author{B. J. Taylor and M. B. Maple}
\affiliation{University of California at San Diego, La Jolla, CA-92093}
\author{}
\affiliation{ }
\date{\today}
\begin{abstract}
         The nature of the out-of-plane dissipation was investigated in
underdoped $Y_{0.54}Pr_{0.46}Ba_{2}Cu_{3}O_{7-\delta}$ single
crystals at temperatures close to the critical temperature. For this
goal, temperature and angle dependent
out-of-plane resistivity measurements were carried out  both below and above
the critical temperature. We found that the Ambegaokar-Halperin
relationship [V. Ambegaokar, and B. I. Halperin, Phys. Rev.
Lett. \textbf{22}, 1364 (1969)] depicts very well the angular
magnetoresistivity in
the investigated range of field and temperature. The main finding is
that the in-plane phase fluctuations decouple the layers above the
critical temperature and the charge transport is governed only by the
quasiparticles.
We also have calculated the interlayer Josephson critical  current
density, which was
found to be much smaller than the one predicted by the theory of
layered superconductors.
This  discrepancy could be a result of the $d$-wave symmetry of the order parameter and/or of the non
BCS temperature dependence of the $c$-axis penetration length.

\end{abstract}
\maketitle

Since the discovery of high temperature superconductivity in cuprates, the
contrasting temperature $T$ dependence of in-plane $\rho_{ab}$ and
out-of-plane $\rho_{c}$ resistivities have been an issue of debate. This
topic is even more complex in the
underdoped systems where the density of states DOS of quasiparticles as well
as the
transfer integrals are momentum and temperature dependent. Additionally,
strong fluctuations, which are expected at low carrier density,
have a major contribution to dissipation.

The most debated issue is the
interlayer dissipation and its field and angle dependence.
In the most general way, the $c$-axis conductivity is presumably
controlled by the
tunneling of the Cooper pairs and quasiparticles, with conductivities
$\sigma_{J}$  and $\sigma_{c,qp}$, respectively; i.e., the $c$-axis
conductivity $\sigma_c =
\sigma_J+\sigma_{c,qp}$. This tunneling is the main consequence of the layered
structure of the cuprates, which can be depicted as stacks of
Josephson junctions made out of superconducting CuO$_2$
"electrodes"
and intermediate blocking layers. Actually, this Josephson coupling
of the layers distinguishes the layered cuprates from the ordinary anisotropic
superconductors.
The existence of this coupling was unambiguously demonstrated for
cuprates with large anisotropy
$\gamma>100$, where $\gamma$ is the ratio of the $c$-axis and
in-plane effective masses,
$\gamma\equiv m_{c}/m_{ab}$, either directly by $I-V$ characteristics
on small single crystals or messa
structures \cite{Kleiner1, Kleiner2, Sakai, Irie, Yurgens, Latyshev}
or by Josephson plasma resonance experiments.
\cite{Matsuda,Tsui, Bulaevski1, Matsuda2, Hanaguri, Gaifulin} In the
case of systems with lower anisotropy, e.g.,
YBa$_2$Cu$_3$O$_{7-\delta}$ ($\gamma\approx 5-9$), the existence of a
Josephson coupling in the $c$-axis direction was largely debated.
Direct measurements
were reported only in underdoped YBa$_2$Cu$_3$O$_{6+x}$.\cite{Rapp}
while optical conductivity  measurements
\cite{ Basov, Homes} showed evidence of Josephson coupling in these
low anisotropic cuprates.

As in the case of the
in-plane transport, the interlayer dissipation is strongly influenced by phase
fluctuations. A magnetic field applied perpendicular to the layers
penetrates as pancake
vortices with a particular phase distribution around each core. It is
known that position fluctuations of the pancake
vortices due to pinning and/or
thermal diffusion of the vortex core, reduce phase correlations,
hence, the Josephson
coupling. Nevertheless, as long as phase correlations exist, they could provide
a
Josephson-type contribution
to the out-of-plane transport. In fact, such phase correlations,
though weak, were
identified  experimentally even in the liquid state of the vortex system.
\cite{Cubitt, Tsui, Matsuda3}
Generally, these phase (vortex) fluctuations have important consequences both
below and above the critical temperature $T_{c0}$. For $T<T_{c0}$,
they drive the vortex
system in a liquid state, whereas above $T_{c0}$, they allow vorticity to
survive and to contribute to the in-plane
dissipation.\cite{Sandu, Katuwal}
To be specific, in the latter case, there is a crossover in the
in-plane dissipation from a regime of pure flux-flow to a regime
entirely due to
quasiparticles, which occurs at a particular temperature {\it higher} than
$T_{c0}$. The importance of the fluctuations increases when the
density of charge carriers is reduced,
i.e., in the case of underdoped cuprates.

Even though the interlayer coupling was investigated on a large extent
in large-$\gamma$  superconductors, the
scarcity of  data is evident
in low and medium-anisotropic superconductors. Therefore, in the
present study, using temperature, field,
and  angle dependence of the out-of-plane resistivity, we investigate the
features of the
$c$-axis dissipation in a medium-$\gamma$ superconductor in an
attempt to find the extent of Josephson response close to
$T_{c0}$, i.e., in a temperature range where the phase fluctuations are
important enough to
reduce and/or suppress the Josephson coupling. As in the case of
in-plane resistivity, \cite {Sandu, Katuwal} we
take advantage of the different angular dependences of the different
contributions to resistivity
to obtain the desired information. This investigation is
performed on Y$_{0.54}$Pr$_{0.46}$Ba$_{2}$Cu$_{3}$O$_{7-\delta}$ single
crystals, where we used the antidoping effect of praseodymium to reduce the
charge carrier density and increase the electronic anisotropy,
hence, to drive the system in the strong fluctuation regime.
In this way the electromagnetic and Josephson coupling of the pancake
vortices
and, subsequently, the interlayer coherence are weakened compared with the
thermal fluctuations.
Our main finding is that the $c$-axis dissipation scales with
$Hcos\theta$ below $T_{c0}$, in a temperature range that ends at the critical
temperature. We relate the failing of the scaling above $T_{c0}$ to the
suppression of the interlayer Josephson coupling by the
in-plane phase fluctuations. So, even though in-plane strong
superconducting fluctuations, hence, a large amount of condensate,
persist above $T_{c0}$,\cite{Sandu, Katuwal} there is no out-of-plane
Josephson coupling
which  would provide an enhanced conductivity above $T_{c0}$.

Y$_{0.54}$Pr$_{0.46}$Ba$_{2}$Cu$_{3}$O$_{7-\delta}$ single crystals
with typical
dimensions of 1.0
$\times$ 0.5 $\times$ 0.02 mm$^{3}$ were grown using a standard procedure
described elsewhere. \cite{Paulius} The dimensions the the crystal 
for which the data are shown
here are 0.6
$\times$ 0.65 $\times$ 0.017 mm$^{3}$. We attached four gold wires 
(0.025 mm in diameter) with
silver epoxy onto each of the two large faces of the single crystal
(see top Inset to Fig. 1). The two
outer (inner)
contacts on the same face were used as current (voltage) terminals. The
contact resistance is 2 $\Omega$ at room temperature. First, a
magnetic field H up to 14 T was applied along the $c-$direction of 
the sample, a
constant current
$I\leq 1$ mA was fed through pads of size 0.05 $\times$ 0.5 mm$^{2}$, 
alternately
on both  faces, and the voltage on each face of the single crystal
was measured at set temperatures between
0 and 300 K. Next, we measured in the same way the voltages at
different constant temperatures but this time
the single crystal was rotated in the applied magnetic field with the
angle $\theta$ between
$H$ and the $c$-axis varying between 0 and 360$^\circ$. The
out-of-plane $\rho_{c}$ and in-plane $\rho_{ab}$ resistivities were
calculated using  an algorithm described elsewhere.\cite{Levin} The critical
temperature $T_{c0}$ was taken at
the midpoint of the normal-superconductor transition.

Figure 1 shows the temperature $T$ dependence of $\rho_c$ of an
Y$_{0.54}$Pr$_{0.46}$Ba$_{2}$Cu$_{3}$O$_{7-\delta}$ single crystal measured at
different applied magnetic fields, while its bottom
Inset shows the zero field $\rho_{c}(T)$ over the whole measured
temperature range. Even though this is a
strongly underdoped sample, with a zero-field  superconducting transition
temperature $T_{c0}=38$ K, it has a sharp transition,
attesting to the good quality of this single crystal. The
magnetoresistivity in the normal
state is very small and positive.  The normal state is metallic at
high temperatures and becomes nonmetallic for
temperatures lower than 133 K. The upturn in $\rho_{c}(T)$ observed below
this temperature is the result of a complex process. First, there is a
reduction in the planar density of states DOS due to the opening of
the pseudogap at
$\mathbf{k}=(\pi/2a,0)$ (nodes) with decreasing $T$;\cite{Damascelli}
second, the transfer
integral of the coherent contribution is angle dependent with maxima
at the nodal points.

The angular dependence of the normalized $c-$axis resistivity
$\rho_{c}(\theta)/\rho_{c}(\theta = 0^\circ)$ of
$Y_{0.54}Pr_{0.46}Ba_{2}Cu_{3}O_{7-\delta}$ measured at 30, 35 and 40
K in a magnetic field of 14 T
is shown in Fig. 2.  Both below and above $T_{c0}$, $\rho_c(\theta)$
displays a minimum at $\theta=90^{\circ}$  (i. e. for
$H\parallel ab-$plane) and a maximum value at
$\theta = 0^\circ$ (i. e. for $H\parallel c-$axis). The former
(latter) value of the
angle corresponds to maximum (zero) transverse Lorentz force on the
flux vortices.
This fact rules out the possibility that flux motion contributes to
the measured
$c-$axis dissipation, since the measured dissipation is maximum for
angles for which
the Lorentz force is zero. Therefore, we assume that the
$c$-axis transport in the mixed state involves Josephson and quasiparticle
contributions which depend on $T$, $H$, and $\theta$.

In the geometry we have used, the $c$-axis component of the current
density flows mainly
along the crystal edges in a wall of width almost equal to the
pad width $\Delta=50$ $\mu$m. We call this region the
active area. The variation of the  current
density over this width  is $[J_{z}(z, L/2)-J_{z}(z,
L/2-\Delta)]/J_{z}(z, L/2) <14$ \%, while, outside of this
width, the current density decreases 
fast toward the  center of the crystal.
Therefore, we conclude that only  these edge walls of the single crystal
are involved in the current transport along c-axis. 

The characteristic
lengths for a  stack of Josephson
junctions are the Josephson penetration length $\lambda_{j}$, which 
accounts for the field
penetration within the nonsuperconducting interlayer space and the 
$c$-axis London penetration
length $\lambda_{c}$, which accounts for the magnetic screening. For this single crystal,
$\lambda_{j}
\approx 0.03$ $\mu$m, much smaller than $\Delta$. Therefore, it is inappropriate to
consider the single crystal as a stack of short junctions. For a stack
of long junctions, Josephson vortices might be generated even in the
absence of in-plane external fields if the in-plane
currents are strong enough to create the required phase gradient. This phase gradient is generated 
if the junction length is larger than $\lambda_{c}$.
Hence, if  the width of the active area $\Delta$ is of the order of the $c$-axis magnetic penetration
length
$\lambda_{c}$ one can assume that the
in-plane currents are too small to generate Josephson vortices in
the active area. There are no available data concerning $\lambda_{c}$ for
strongly underdoped Y$_{1-x}$Pr$_{x}$Ba$_{2}$Cu$_{3}$O$_{7-\delta}$. 
An estimate
of
$\lambda_{c}$ based on a similar underdoped YBa$_2$Cu$_3$O$_{x}$ is of the
order of 30 $\mu$m.
\cite{Hosseini}  Therefore, this Josephson system satisfies the condition
$\lambda_{j}
\ll\Delta \sim\lambda_{c}$. Hence, we assume that in this case the 
Josephson vortices are most likely
absent and we can use the Ambegaokar-Halperin relationship. 
Additionally, the current we use
is very low with a total current density $J=4$ Acm$^{-2} \ll J_{c}$. 
Under these assumptions, we proceed to obtain
an analytical relationship for the $c$-axis resistivity
as a function of field, temperature, and angle.

Generally, the total conductivity can be derived from Kubo's
relationship for both Josephson
and quasiparticle current.\cite{Koshelev2} However, this
relationship is difficult to handle in the absence of an analytical
dependence of the in-plane diffusion coefficient on field
and temperature. Several experimental
reports\cite{Briceno,Gray,Hettinger,Yoo} have
shown that the Ambegaokar-Halperin (AH) expression, \cite{Ambegaokar}
which is valid
for a single Josephson junction, can be successfully used to fit the
$c$-axis resistivity
data when making specific assumptions on the expression of the
critical current.
The  Ambegaokar-Halperin expression for $c$-axis resistivity
is given by:

\begin{equation}\rho_{c}(T) = \rho_{n} \left[\mathcal{I}_{0}\left(\frac
{\Phi_0
I_{c}(T)}{2\pi k_{B} T}\right)\right]^{-2},
\end {equation}
where $\rho_n$ is the intrinsic normal-state resistivity of the
junction, $\mathcal{I}_{0}$ is
the  modified Bessel function, $\Phi_0$ is the flux quantum,
$I_{c}$ is the critical current at a temperature T, and
$k_{B}$ is Boltzmann's constant.
Note that the AH relationship accounts also for the contribution of the
quasiparticles through $\rho_n$.

Because of the high energy of the
Josephson coupling $\Phi_0 I_{c}/2\pi$ relative to the thermal
energy, one can use the asymptotic expansion
$\mathcal{I}_0(x)\approx
\exp(x)/\sqrt{2\pi x}$ for $x > 1$. This approximation gives, for example, for
$x=3$ a 5\% error compared
with the exact Bessel function. An estimate of the zero-field conductivity of
our samples gives $x(B=0)=5.5$ at 35 K and an error of approximately 2\%.
With this approximation, Eq. (1) gives the the following expression
for the $c-$axis
resistivity at high temperatures:

\begin{equation}
\rho_{c}(T,H)\approx
\rho_{c,qp}\frac{\Phi_0I_{c}(T,H)}{k_{B}T}{\exp\left(-\frac{\Phi_0I_{c}(T,H)}{\pi 
k_{B}T}\right)}.
\end{equation}

Next, we discuss the $T$ and $H$ dependence of $I_c$. The temperature
dependence of the
AH  relationship
is limited only to the spin-wave type fluctuation of the order parameter.
Therefore, we have to include in the above critical current term the
contribution
accounting for the presence of vortices. The maximum  Josephson current
$I_{c}$, which is related to the interlayer phase difference, is strongly
influenced by the level of the fluctuations  of the phases in each layer.
To be specific, the Josephson current density $J_{c}$ decreases as a result of
both the thermally-induced misalignment of the planar
vortices, which creates a gauge invariant phase difference
$\varphi_{n,n+1}$ from layer to layer, and the thermally induced phase
slippage (spin-wave type phase fluctuations).
Regarding the thermal motion of the pancake vortices, there is a
complex process of renormalization of $J_{c}$,
which suppresses $J_{c}$.\cite {Daemen} A suppressed $J_{c}$ increases
in turn the penetration depth
$\lambda_{c}$, hence, reduces the elastic constants, which in turn
further suppresses $J_{c}$. This process is present both
in the solid and liquid phases of the vortex system because
the only difference between these two phases is the vanishing of the shear
constant $c_{66}$ in the latter
phase. Additionally, the phase difference depends on the pancake
position within each plane. Therefore, to obtain the interlayer critical
current, one has to go beyond the ensemble average
\cite{Koshelev4, Koshelev3} and to  also use a space average of the
critical current. Hence, \cite{Fistul,Logvenov}
\begin{equation}
I_{c}^2=J_{c0}^2\!\int\!\!d\mathbf{r}_1\!\!\int\!\!d\mathbf{r}_2
\exp\left
\{i\left[\phi_{n,n+1}\left(\mathbf{r}_1\right)-\phi_{n,n+1}\left(\mathbf{r}_2\right)\right]\right\},
\end{equation}
with $J_{c0}$ the local intrinsic (bare) Josephson critical current
density. At high temperatures, Eq. (3) becomes:
\begin{displaymath} I_{c}^2 (T, H)=J_{c0}^2(T)AS(H),
\end{displaymath} where $A$ is the
area of the junction and $S(H)$, given by

\begin{equation} S(H)=\int d\mathbf{r}\langle
cos\left[\phi_{n,n+1}\left(\mathbf{r}\right)\right]-cos\left[\phi_{n,n+1}(0)\right]\rangle,
\end{equation}
is the correlation area, which needs to be evaluated. Following
Koshelev, Bulaevski, and Maley, \cite{Koshelev3} we
make the approximation
\begin{equation}
S(H)\approx f\gamma^2s^2\left(H_{j}/H_{z}\right)^\alpha.
\end{equation}
        Here,
$f(H,T)$ is a function of order  unity with a weak $T$ dependence,
$s$ is the interlayer spacing, $H_{j}=\Phi_0/\gamma^2
s^2$ is a characteristic field, and $H_{z}=Hcos\theta$ is the
magnetic field in the
$c$-axis direction. The deviation of the exponent
$\alpha=1-k_{B}T/2\pi E_0(T)$ from unity
is a result of the spin-wave type phase fluctuations and increases
strongly close to the critical
temperature. [$E_0=s\Phi_{0}^2/(4\pi\mu_0\lambda_{ab}^2)$ is the
Josephson energy of the area $\gamma^2 s^2$].\cite{Koshelev1}
The above approximation [Eq. (5)] is
valid for applied magnetic fields larger than the
characteristic field $H_{j}$.
For the single crystals
with a Pr doping $x$ = 0.46, for
which
$\gamma=26$ and $s=11.7 \AA$, the characteristic field $H_{j}\approx 2.3$ T.
Hence, Eq. (5) is valid for $H>2.3 $ T. With this
approximation,
the Josephson critical current becomes

\begin{equation}
I_{c}(H,T)=J_{c0}(T)\gamma s f^{1/2}A^{1/2}(H_j/H_{z})^{\alpha/2}.
\end{equation}
Equation (4), hence Eq. (6), was derived in the approximation of a completely
decoupled  pancake vortex liquid, when the correlation of the
$cos\varphi_{n,n+1}$ terms drops on a length scale of the order of the
intervortex
spacing.

Equations
(2) and (6) give the following expression for the out-of-plane
resistivity:
\begin{displaymath} \rho_{c}(T,H,\theta)\approx
\rho_{c,qp}\frac{\Phi_0J_{c0}(T)\gamma s
f^{1/2}A^{1/2}}{k_{B}T}\left(\frac{H_{j}}{H|cos\theta|}
\right)^{\nu}\times\end{displaymath}
\begin{equation}\;\;\; \exp\left [-\frac{\Phi_0J_{c0}(T)\gamma s
f^{1/2}A^{1/2}}{\pi k_{B}T}\left(\frac{H_{j}}{H|cos\theta|}
\right)^{\nu}\right], \end{equation} with $\nu=\alpha/2$. An
important result of Eq.
(7) is that the out-of-plane resistivity in the mixed state where Josephson
tunneling dominates the conduction should scale with
$H\cos\theta$.
Figure 3(a) is a plot of $\rho_c$ vs $Hcos\theta$, measured at several
temperatures below
$T_{c0}$ and in applied magnetic fields of 6, 8, 10, 12, and 14 T. Note
that the data, indeed, follow the $H\cos\theta$ scaling.

Above the critical temperature, $\rho_{c}$ vs $H
cos\theta$ plots
do not map anymore onto a single curve [see Inset to Fig. 3(a) for $T$ = 40 K].
This fact hints to the complete vanishing of the interplane phase
correlations, hence, of the Josephson contribution to dissipation
above $T_{c0}$. Thus, the phase fluctuations suppress the mechanism
responsible for bulk superconductivity at the critical
temperature, even though the in-plane dissipative processes still carry
the hallmark of superconducting phase fluctuations
up to temperatures well above $T_{c0}$.\cite{Sandu, Katuwal}

The term $\rho_{c,qp}$ in Eq. (7) ensues from the quasiparticle
current driven by the time variation of the gauge invariant phase
difference. In the case of d-wave superconductors, the quasiparticle
concentration does not vanish with decreasing temperature. At any
temperature $T$, there is always a $\mathbf {k}$ range so
that $\Delta (\mathbf{k})< k_{B}T$, which facilitates the quasiparticle
excitation near the gap nodes. Microscopic models
have shown that in the case of constant DOS, the quasiparticle out-of-plane
resistivity below the critical temperature depends on
temperature as $\rho_{c,qp} = \rho_{n} (3 \Delta_{0}^{2}/\pi T^2)$ if the
tunneling is coherent and is $T$-independent if the
tunneling is completely incoherent.\cite{Artemenko} A real material
displays both contributions, hence, it follows a power law
temperature dependence.
Additionally, the DOS decreases with decreasing $T$ in underdoped cuprates. The
absence of an analytical expression for the temperature dependence of
the DOS makes impossible the
determination of the
temperature dependence of the quasiparticle contribution to
resistivity.
In the presence of a magnetic field, there is a small
change in conductivity due to the Doppler
shift in the quasiparticle spectrum.\cite{Vekhter}
However, a
sensitive change requires extremely high magnetic
fields,\cite{Morozov} so that in
the present measurements ($H\leq 14 $ T)
$\rho_{c, qp}$ is practically field independent.  Indeed, the
$c$-axis resistivity data show that the normal-state magnetoresistivity is very
small, which implies an almost $H$ independent quasiparticle contribution
to the
out-of-plane conduction. Therefore, we assume that $\rho_c(T)$ measured in 14 T
in the normal state and its extrapolation at lower temperatures in
the mixed state is the
out-of-plane quasiparticle resistivity $\rho_{c,qp}(T)$.

With $\rho_{c,qp}(T)$ determined as just discussed above, we fit the
$\rho_c(T,H,\theta)$ data with Eq. (7) with two fitting parameters:
the exponent
$\nu(T)$ and

\begin{displaymath}C(T)=\frac{\Phi_0 J_{c0}(T)\gamma s
A^{1/2}H_{j}^{\nu(T)}}{\pi k_{B}T},\end{displaymath} where we take
$f\approx 1$. The results of the fitting of the data measured at several
temperatures, and in 14 T and 10 T are shown in Fig. 3(b) and its Inset,
respectively. The excellent fit of the out-of-plane resistivity data
with Eq. (7)
confirms the validity of our approach and shows that the $T$,
$H$, and
$\theta$ dependence of the measured out-of-plane
resistivity in the mixed state is dominated by the Josephson tunneling of the
Cooper pairs and the quasiparticle tunneling. 

The picture that immerges from these results is as follows. As in the case of
the in-plane dissipation, the fluctuations have a significant
effect on the nature of $\rho_{c}$ at high temperatures in low charge
carrier density cuprates such as
Y$_{0.54}$Pr$_{0.46}$Ba$_{2}$Cu$_{3}$O$_{7-\delta}$. Nevertheless,
although the dissipations along the two directions have the same
origin, they are governed by different mechanisms, hence, display two
temperature scales.
The out-of-plane dissipation is
governed by the Josephson tunneling of the Cooper pairs and the
quasiparticle tunneling.
With increasing $T$, the in-plane phase fluctuations give rise to a
rapid suppression
of the Josephson coupling, the unique process which makes
superconductivity a bulk phenomenon, and of the corresponding
interlayer supercurrent density.
Small interlayer correlations survive above the
irreversibility temperature up to 37 K above which both the Josephson
coupling and the corresponding  interlayer supercurrent vanish.  At $T>T_{c0}$,
even though in-plane vorticity has been shown to exist in a strong
fluctuating regime, \cite{Sandu} these phase fluctuations are too
fast to allow any phase
correlation  along the $c$-axis. Hence, $\sigma_{J}=0$ and the
out-of-plane conductivity is only a result of quasiparticle
tunneling. This is not
the case of  the in-plane
dissipation, which displays a
contribution from phase fluctuations (pancake vortices) arising from
their motion driven by the
transport current up to a charge carrier density dependent temperature
$T_{\varphi}>>T_{c0}$.\cite{Sandu, Katuwal} Therefore, the temperature
scale is $T_{c0}$ ($T_{\varphi}$) for
the contribution of the superconducting dissipation to the total out-of-plane
(in-plane) resistivity.
These two temperatures merge as the density of charge
carriers increases toward the optimal doping.

Additional improvements to the model used in this study would require
the incorporation
of nonequilibrium effects due to the nodal quasiparticles,
mainly in the high temperature range.\cite{Artemenko2}
This simple AH approach, which constitutes the starting point of our
analysis, though fruitful, has been the subject of  criticisms regarding
the omission of the interaction
between adjacent junctions, \cite{Goldobin} and the assumption of a Fermi
liquid behavior in underdoped cuprates.
However, the simplicity of the AH relationship and
its reported success
at low temperatures make it attractive, with appropriate assumptions,
in the high
temperature regime. Our present results confirm, indeed, its
applicability close to
$T_{c0}$.

 From the angular
magnetoresistivity data, we also extracted the temperature dependence of the
bare interlayer (Josephson) critical current density at high 
temperatures, close to
$T_{c0}$, from
\begin{equation} J_{c0}(T) = \frac{\pi
k_{B}TC(T)}{\Phi_0\gamma s A^{1/2}H_{j}^{\nu(T)}},
\end{equation}
using the fitting parameter $C(T)$ as obtained from the
fit of the angular magnetoresistivity data at high temperatures with Eq. (7). A
plot of $J_{c0}$ vs $T$ is shown in Fig. 4.
These values of $J_{c0}$ are
smaller than the values smaller  than the values predicted by a simple model of layered
superconductors,
which gives
$J_{c0}=\Phi_0/[2\pi \mu_0\lambda_{c}^2(T)]$. 
The temperature dependence of the
data follow the power law
$J_{c0}(T)\approx 7.6 (T/T_{c0})^{-1.73}$. Such a $T$ dependence could
be the result of the complexity of the interlayer Cooper pair
transport in cuprates containing conducting CuO chains combined with
the $d-$wave symmetry of the superconducting order
parameter. Therefore, the temperature dependence of $J_{c0}$ is
provided not only by $\lambda_{c}^{-2}(T)$, but also by the
$T$-dependence of the density of states of the localized resonant centers.\cite{Abrikosov}
The $\lambda_{c}^{-2}(T)$ itself changes its convexity at high
temperatures, \cite{Hosseini} most probably due to the excitation of
the quasiparticles out of the condensate at gap nodes.

The temperature dependence of the exponent $\nu$ is shown in the Inset to
Fig. 4. A fit of the data gives a linear $T$ dependence, i.e.,
$\nu(T)=1.7(1-T/T^{*})$, where $T^{*}=39.5$ K is slightly higher than
$T_{c0}$ defined
as the mid point of the transition curve. Theoretically, the exponent
$\nu$ should be linear in
$T\lambda_{ab}^2(T)$.  A plot of the theoretical $\nu(T)$ curve using the BCS
dependence of
$\lambda_{ab}^2(T)$ is also shown in Fig. 4. The experimental and
theoretical values are close to each other for $T\geq 30$ K. At lower
temperatures, the data obtained from fitting are almost twice as high as the
theoretical values. Actually, the expression used for the field
dependence of $I_{c}$ [Eq. (6)] is
not valid at low temperatures. This could be one reason for the above
discrepancy. It is interesting to note, however, that the
scaling of
$\rho_{c}(H|cos\theta|)$ still works down to 20 K.   Another reason for
the discrepancy between the experimental and
theoretical values of $\nu$ could be that $\lambda_{ab}^{-2}(T)$ has a non BCS
temperature dependence due to the nodal quasiparticles.

In summary, we analyzed the out-of-plane dissipation in a medium
anisotropic underdoped cuprate at temperatures around $T_{c0}$. We
performed these measurements in order to investigate the origin of
the large $\rho_{c}$ and its $T$, $H$, and angle dependence in
this material. The data are well fitted by the Ambegaokar-Halperin 
expression for
temperatures up to the critical temperature and applied magnetic fields as
high as 14 T. We found that the interlayer resistivity follows a
simple scaling law as a function of magnetic field and angle; i.e.,
$\rho_{c}(H,\theta)=
\rho_{c}(|Hcos\theta|)$. The existence of the scaling close to the critical
temperature proves the persistence of
interlayer correlations above the irreversibility temperature.
Nevertheless, the scaling fails above the mid point critical
temperature, above which the $c$-axis charge transport is governed by
quasiparticles only. This is different from the in-plane dissipation,
in which the contribution of the superconducting fluctuations can be
discerned up to temperatures as high as
$1.5\times T_{c0}$. We also have determined the interlayer critical current
density. It was found to be lower than predicted by simple
models of Josephson coupled superconductors.

\textbf{Acknowledgments}

This research was supported by the National Science
Foundation under Grant
No. DMR-0406471 at KSU and the US Department of Energy under Grant No.
DE-FG02-04ER46105 at UCSD. \label{}

\newpage
\begin{figure}[tbp]
\caption{Temperature $T$ dependence of out-of-plane resistivity $\rho_{c}$
of an Y$_{0.54}$Pr$_{0.46}$Ba$_{2}$Cu$_{3}$O$_{7-\delta}$ single
crystal, measured in an
applied magnetic field of 0, 6, 8, 10, 12, and 14 T and for $T\leq
80$ K. Insets: (top) Sketch of sample geometry and leads configuration.
(bottom) Zero field
$\rho_c(T)$ shown over the whole measured $T$ range. The solid lines
are guides to
the eye.}
\end{figure}

\begin{figure}[tbp]
\caption{Angular $\theta$ dependence of normalized out-of-plane
resistivity $\rho(\theta)/\rho(0)$ of an
Y$_{0.54}$Pr$_{0.46}$Ba$_{2}$Cu$_{3}$O$_{7-\delta}$ single crystal,
measured at 30, 35,
and 40 K and 14 T.}
\end{figure}

\begin{figure}[tbp]
\caption{(a) Plot of the out-of-plane resistivity $\rho_c$ vs
$H|cos\theta|$ of an Y$_{0.54}$Pr$_{0.46}$Ba$_{2}$Cu$_{3}$O$_{7-\delta}$
single crystal,
measured at 25, 30, 35, and 36 K and 6 T (open circles), 8 T (diamonds), 10 T (triangles), 12 T
(inverted triangles), and 14 T (open squares). Inset: Plot of
$\rho_c(H|\cos\theta|)$ measured at 40 K. (b) and its Inset: Same
plot of the data measured at 14 T
and 10 T, respectively. The solid lines are fits of the data with Eq. (7).}
\end{figure}

\begin{figure} [tbp]
\caption{Josephson critical current density $J_{c0}$, calculated with the
fitting parameters
obtained by fitting the angular
magnetoresistivity, vs reduced
temperature $T/T_{c0}$. The solid line is a power-law fit. Inset:
Temperature $T$ dependence of exponent $\nu$ (empty circles)
obtained by fitting the  magnetoresistivity data. The theoretical $\nu(T)$
dependence is calculated with the $T$ dependence of the penetration depth
$\lambda_{ab}(T)$ given by the BCS theory and taking as the critical
temperature  $T_{c0}=38$ K.}
\end{figure}
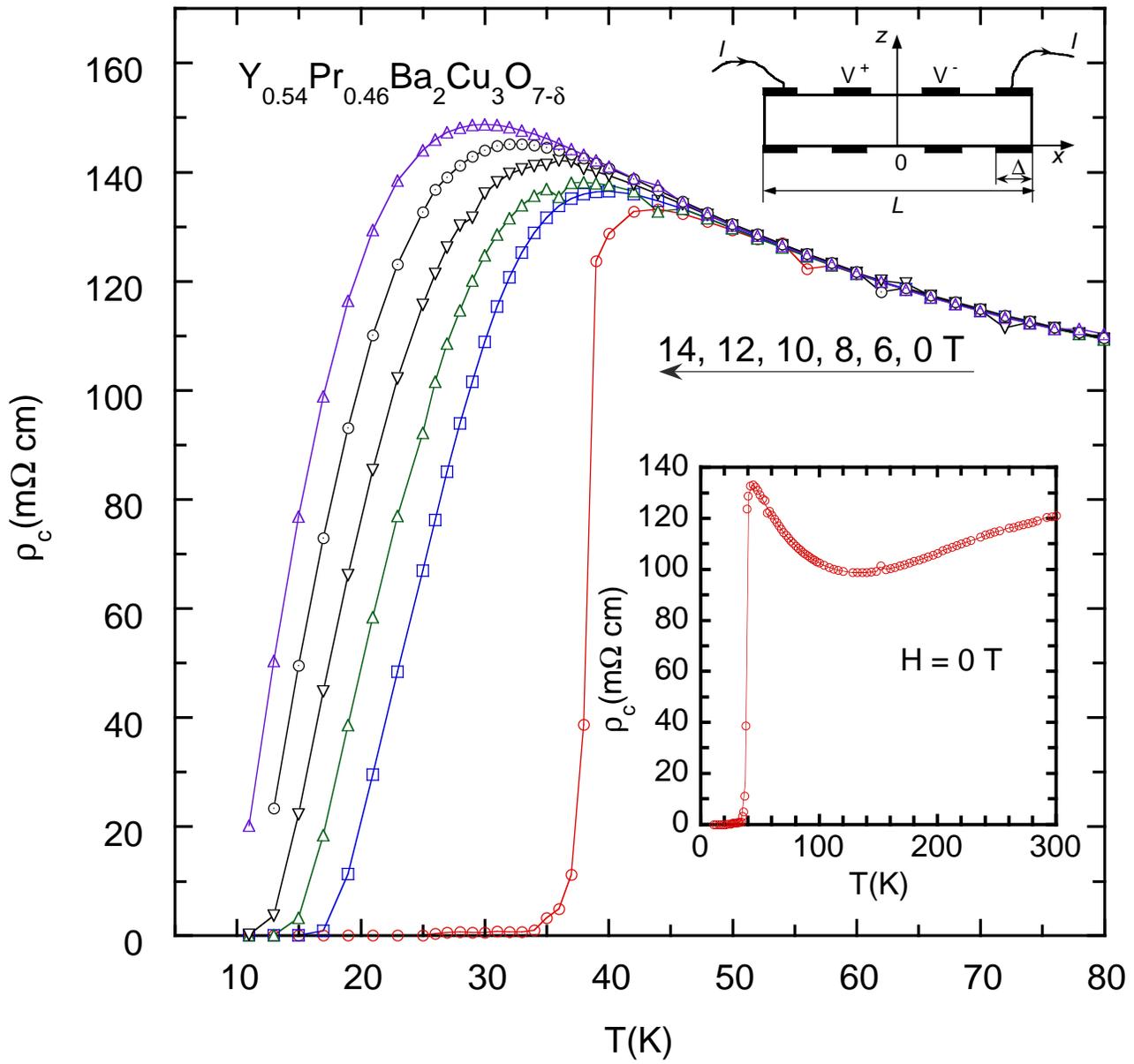
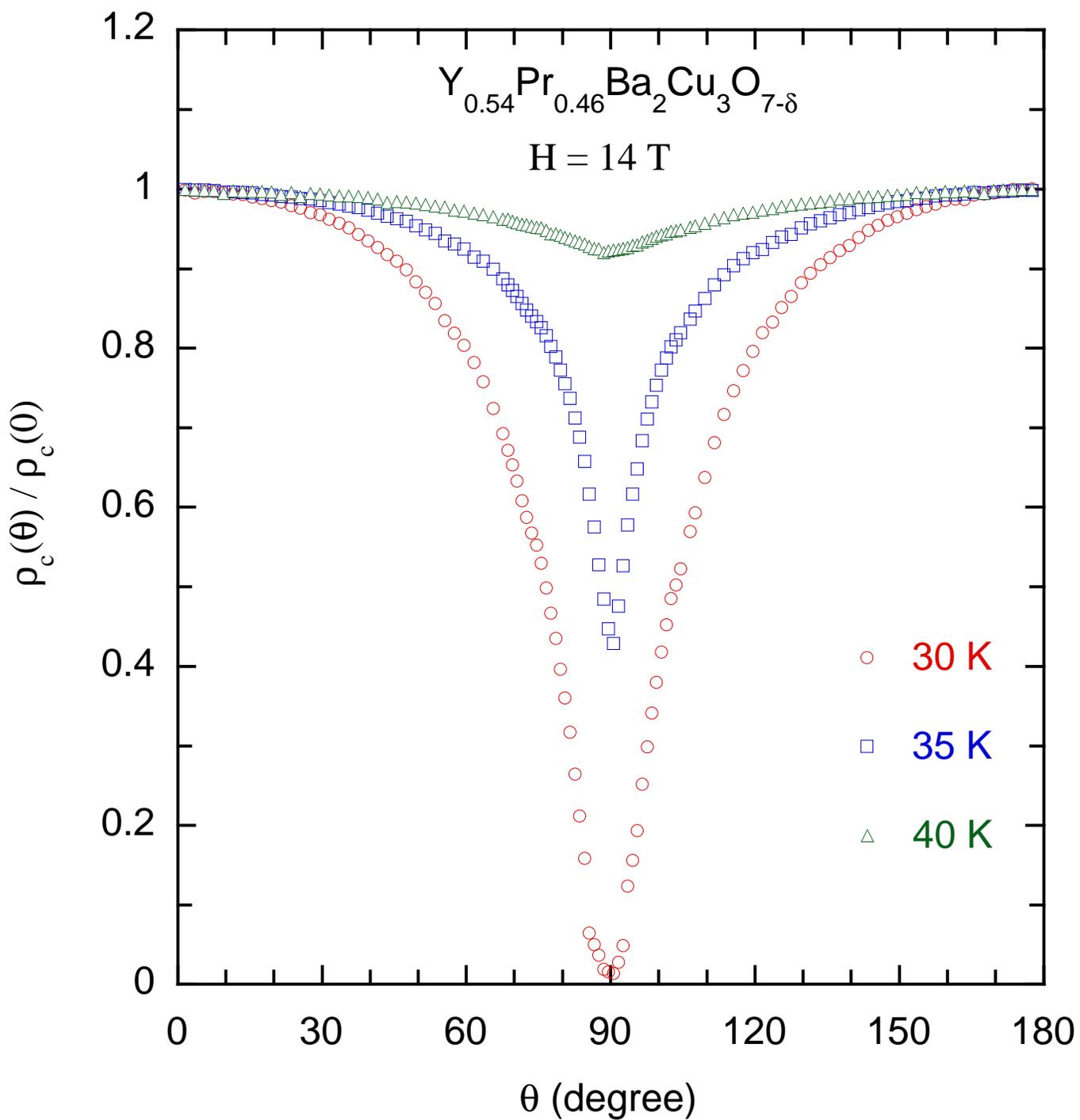
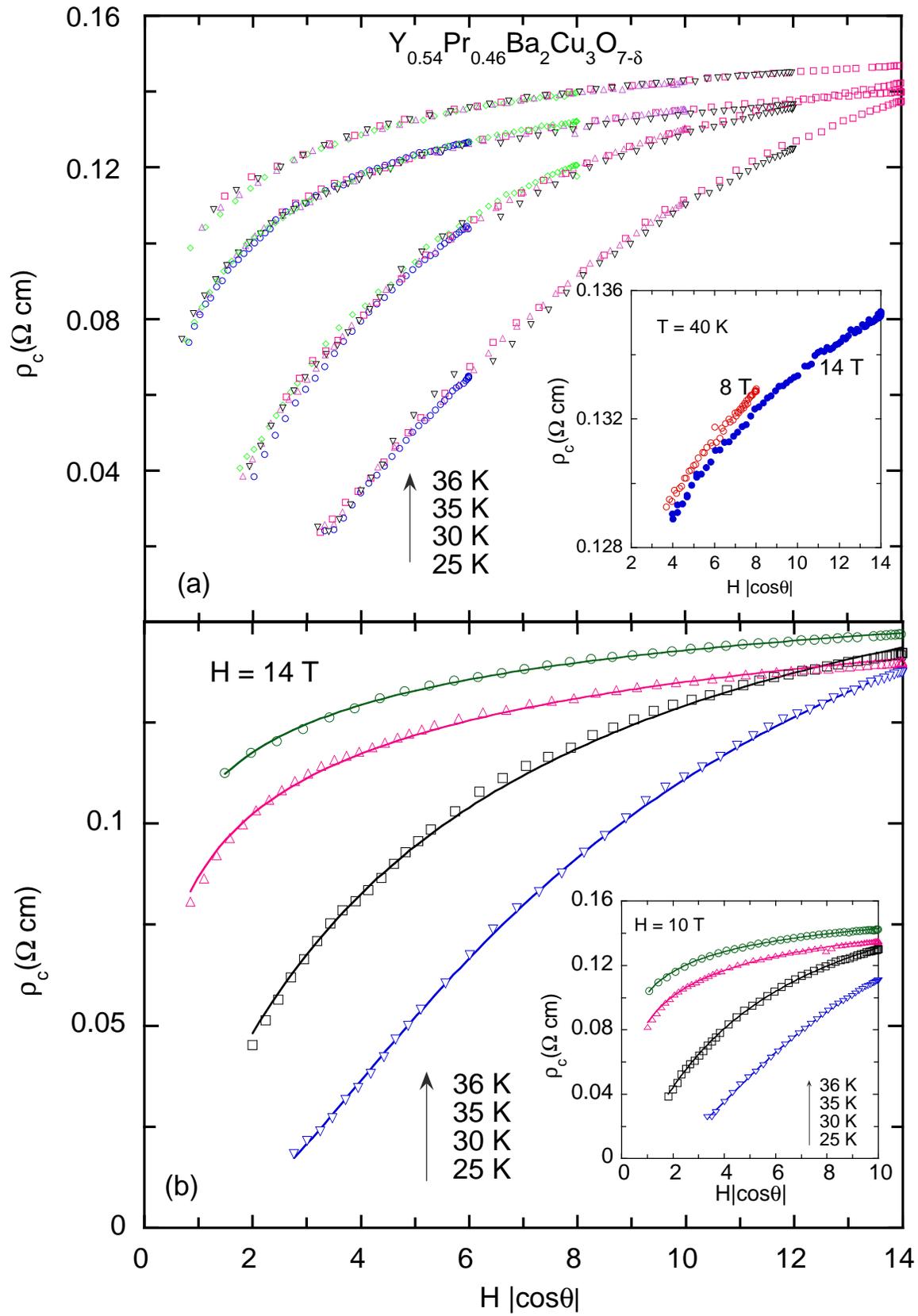
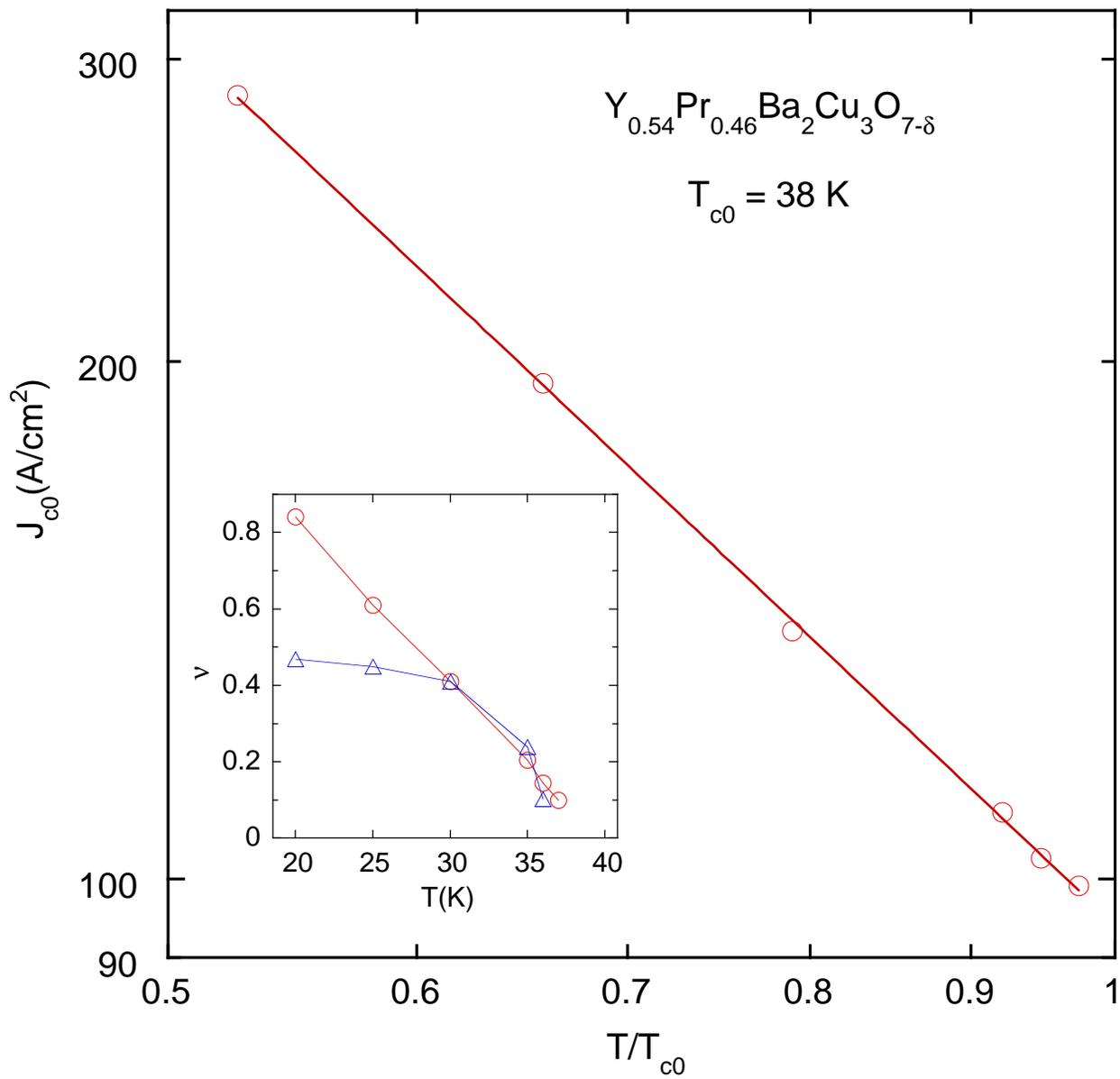
\end{document}